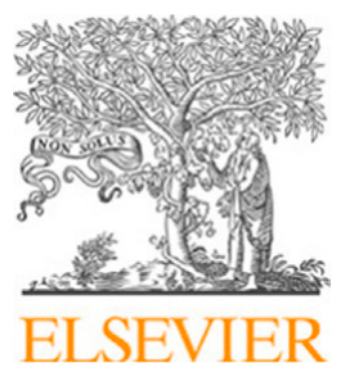

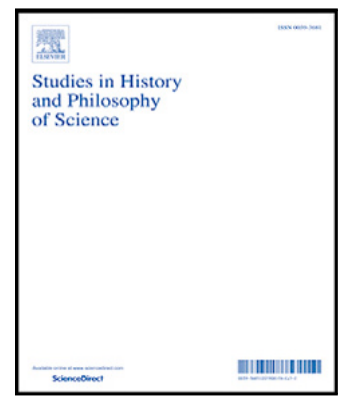

# The Dynamics of Quantum Gravity: The Missing Piece in the Spacetime Emergentist Account

Álvaro Mozota Frauca

*Department of Architectural Technology, Division of Mathematics, Polytechnic University of Catalonia, Avinguda Diagonal 649, Barcelona, 08028, Barcelona, Spain*

ABSTRACT

Quantum gravity suggests that spacetime may not be fundamental, and it has been argued that we can understand a theory without a fundamental spacetime if we are able to claim that spacetime 'emerges' from some non-spatiotemporal entities. In this sense, strategies like functionalism have been deployed to claim that this emergence is possible and plausible, both in principle and in practice for current approaches to quantum gravity. In this article I argue that this analysis is incomplete, as it tends to overlook the way the dynamics of these theories is 'quantum' in a way that differs from standard quantum theory. The challenge for the emergentist (and for the quantum gravity theorist) is to give an interpretation not only to the kinematical and classical aspects of these theories, but to the dynamical and quantum ones, and to show how the spacetime roles can be fulfilled, if possible at all. Therefore, I argue that some current approaches to quantum gravity seem to fail to provide meaningful theories, that spacetime functionalism is of no help, and that the position of the spacetime emergentists is weakened, as they lack any example of a successful reduction of spacetime to some truly quantum non-spatiotemporal stuff.

## 1. Introduction

Some theories of quantum gravity have claimed that they describe a spacetimeless reality. This is conceptually challenging, but it has been argued that we can make sense of these theories at least if they are able to recover a spacetime at some emergent level. Some authors have argued that this emergence is both possible and plausible, not just in principle, but for some of our current approaches. However, in this paper I will argue that this kind of analysis suffers from a fundamental shortcoming, as it typically does not consider a complete theory of quantum gravity and leaves out the quantum and dynamical aspects of these theories. This emphasis on the quantum and dynamical aspects is crucial for my argument, as in many approaches to quantum gravity there is an important departure from the standard quantum formalism and dynamics which renders the usual interpretations and ways of thinking about quantum theory inapplicable. For this reason, I argue that the spacetime emergentist or the quantum gravity theoretician needs to clarify what the candidate theory claims to be there at the fundamental level in order to argue that it is plausible to claim that spacetime emerges from it.

An important clarification to be made is that although there is no consensus about which approach to quantum gravity is on the right track and that quantum gravity is widely considered to be a work in progress, physicists working on different approaches believe they have a more or less developed idea about what a theory of quantum gravity would look like and they have developed different formalisms for which one would require only filling in some technical details. In this sense, we find textbooks like (Rovelli, 2004; Rovelli & Vidotto, 2015) in which theories (canonical and covariant loop quantum gravity) are presented as complete and meaningful, or presentations like the one in Oriti (2017), in which a formalism for quantum gravity is introduced as still flexible and raising some conceptual and philosophical questions, but nevertheless complete in the sense that, if we were to specify a number of technical details, the models allowed by the theory would be well-defined. This means that they take the quantum dynamics they define to be physically meaningful. Spacetime emergentists argue that we can make sense not only of a possible complete theory of quantum gravity that we may develop in the future, but also of some of our approaches as they stand now, including quantum dynamics as is currently proposed. My position in this article is that these ways of proposing dynamics are problematic, which raises worries both about the viability of these approaches and the claim of spacetime emergence in them. From a more general perspective, my arguments weaken the position of the spacetime emergentist. If the putative examples were convincing, they would strongly support emergentism. However, my conclusion that they are not leaves room for skepticism about spacetime emergence.

Let me make an important second clarification. I will be using terms like 'emergence', 'emergent spacetime', or 'spacetime emergentists' to

*E-mail address:* alvaro.mozota@upc.edu.

make reference to the alleged relation between the fundamental, non-spatiotemporal structures postulated by the theories of quantum gravity and the non-fundamental spacetime. With this choice of term I am following the literature on the topic, but it is important to notice that the use of the term 'emergence' does not imply an anti-reductionist attitude. On the contrary, I believe that the way the term is used by the majority of the community is perfectly compatible with a reductionist view in which spacetime would be reduced to the non-spatiotemporal entities proposed by quantum gravity.[1] Thus, my claim that the emergence of spacetime fails can be read as claiming that the ways it has been proposed to reduce spacetime to non-spatiotemporal entities have failed. However, let me also state that this paper is not concerned with the exact metaphysical relation between spacetime and the non-spatiotemporal stuff, and that the worries raised here are quite independent of the way this relation is characterized.[2]

My criticisms in this article have both a technical and an interpretational side. The technical side of the issue has to do with the mathematical structures defined by the approaches considered, which differ significantly from the structures of standard quantum theories. The interpretational side has to do with how to interpret and understand these structures. My argument in a nutshell is that we lack satisfactory interpretations, and there are good reasons to believe that they cannot be provided. Hence, the viability of the approaches to quantum gravity based on these mathematical structures is jeopardized.

I will start this paper by briefly introducing quantum gravity theorizing as a context where spacetime is argued to be lost, and spacetime functionalism as the most promising avenue to recover it. In particular, I will introduce spacetime functionalism as has been defended by Lam and Wüthrich (2018, 2021) in Section 2 and I will also consider the case studies they and other authors propose of how one can recover spacetime in a quantum gravitational context in Section 3. These case studies are based on causal set theory and loop quantum gravity, but I will note that they are limited to 'classical' aspects of the models, that is, even if they include some quantum properties such as discreteness, they fail to give an account of the dynamical and quantum aspects of the models under consideration. In my opinion, it is in these aspects where the biggest challenges for the understanding of these theories lie, and hence where the biggest challenges for functionalism and emergentism in general lie. I will also introduce the analysis of the same case studies by Le Bihan and Linnemann (2019) as a clear example of an analysis that is focused just on the 'classical' aspects of the models considered and that overlooks the many conceptual issues associated with the dynamical and quantum aspects of quantum gravity.

In the next part of the paper (Section 4) I will argue that the departure from the quantum formalism and dynamics of certain approaches to quantum gravity represents a challenge for their interpretation which ultimately jeopardizes the spacetime emergentist project. Therefore, these approaches seem to fail to constitute meaningful physical theories, and ideas like functionalism seem to fail to help us make sense of them. In particular, I will discuss canonical and covariant approaches and I will argue that in both cases there are serious issues at the time of interpreting their formalisms. For this reason, I will argue that they do not seem to describe something able to fulfill spacetime functions or something out of which we could reasonably say that spacetime emerges.

I will finish in Section 5 by reemphasizing the conclusions reached in this article. Namely, when one considers the quantum dynamics, there are serious issues in the interpretation of our current approaches to quantum gravity which seem generic to at least two big families of spacetimeless approaches. As far as I know, spacetime functionalists and emergentists have left aside or underestimated these issues, which to me seem to imply that these approaches do not constitute meaningful theories and that spacetime functionalism and emergentism fail in these cases. Moreover, the general case for spacetime emergence in the case of quantum gravity is weakened severely once one accepts that we lack any example of how a possible spacetimeless theory of quantum gravity could explain our apparently spatiotemporal reality.

## 2. Spacetime emergence for understanding quantum gravity

Different approaches to quantum gravity postulate, for different reasons, that spacetime is not fundamental. That is, according to these theories, there is some set of fundamental degrees of freedom that are not spatiotemporal and which constitute what we know as spacetime. For this reason, one needs to resort to some idea like functionalism as a way of making sense of a world that could be fundamentally spacetimeless and which would have spacetime just at an emergent level. In this section I will give a brief overview of spacetime functionalism[3] as has been proposed in this context in Lam and Wüthrich (2018, 2021)[4] and I will also introduce (Le Bihan & Linnemann, 2019) as an example of an emergentist analysis that is optimistic about our capability for bridging the gap between the spacetimeless and spacetime to make sense of current approaches to quantum gravity.

Several approaches to quantum gravity, most notably loop quantum gravity (LQG) and string theory, posit, argue, or derive that spacetime is not fundamental. Different approaches present different arguments for supporting such a claim. For instance, in LQG the basic structures obtained by quantizing general relativity, loops and spinfoams, are quite different from a continuum space or spacetime. In string theory there are some dualities, i.e., some equivalence relations between theories defined on different spacetimes, that suggest that spacetime may not be fundamental. It is in this context that the question of whether a spacetimeless theory can be a meaningful theory is relevant for understanding how our apparently spatiotemporal world arises.

In the physics and philosophy literature some challenges to this possibility have been raised, taking some debates in the foundations of quantum mechanics as an inspiration. Some views in quantum mechanics take the world to be described by wavefunctions living in configuration space and not by 'local beables', that is, matter living just in 3-dimensional space. Similar arguments to the ones raised against these views have been raised for instance in Esfeld (2021) and Lam and Esfeld (2013) but applied to the case of spacetimeless theories of quantum gravity. A reply to some of these objections can be found in Huggett and Wüthrich (2013).

Spacetime functionalism, in this context, can be seen as a way of answering this kind of objection by trying to articulate the way in which a theory without a fundamental spacetime could describe a world like ours. The way spacetime functionalism is supposed to work is analogous to the way functionalism explains the success of the special sciences. Paradigmatic examples of this are the functional reduction of 'water' to '$H_2O$ molecules' or the functional analysis of the kinetic theory of gases which explains macroscopic properties like pressure or temperature in terms of the motion of molecules. In the case of spacetime functionalism we would have non-spatiotemporal entities organized in such a way that they can fulfill the roles that spacetime plays in our theories.

More explicitly, Lam and Wüthrich characterize spacetime functionalism as:

[1] See Crowther (2016, 2021) for philosophical discussions of which is the best way of understanding the use of the term 'emergence' in the context of quantum gravity.

[2] For the metaphysically inclined reader, I refer the reader to Le Bihan (2018a, 2018b, 2021) and Martens (2019) for works in which the metaphysics of emergent spacetime is discussed.

[3] In the context of the analysis of spacetime theories in more broad terms a different sort of spacetime functionalism has been championed by Knox (2013, 2014). It is beyond the scope of this paper to compare and relate both kinds of spacetime functionalisms.

[4] See also the recent book (Huggett & Wüthrich, 2025), where Huggett and Wüthrich defend the same kind of spacetime functionalism.

> The central tenet of spacetime functionalism is that spacetime, qua ontological posit of a theory, is dispensable if the structures postulated by that theory are sufficient to execute all the relevant functions which would otherwise be performed by spacetime itself. Without pretending to be comprehensive, these functions certainly include the determination of spatial distances and temporal durations, and of spatial and temporal relations between physical objects more generally, and so of their relative localization, i.e., an object's or event's localization in space and time relative to other, usually nearby, objects and events, which thus furnish a local frame of reference. (Lam & Wüthrich, 2021, p. 2)

I believe that for spacetime functionalism to be successful in bridging the gap between a fundamentally spacetimeless theory and our apparently spatiotemporal reality it is a necessary condition to show that this list of functions, or some other very similar to it,[5] can be fulfilled by the fundamental structures of such a theory. The worry I will raise in this paper is precisely that, to the best of my knowledge, we lack any theory of quantum gravity for which we are able to argue that the spacetime functions are fulfilled once we take into account the quantum and dynamical aspects of these theories. I believe that the case studies used for supporting the case of the functionalist rely just on kinematical and classical aspects of the models considered, which is a clear limitation of such analysis.

In this article, I will take the analysis in Le Bihan and Linnemann (2019) as an example of limited emergentist analysis which overlooks the quantum and dynamical aspects of quantum gravity at the time of evaluating the viability of both the claim that spacetime is emergent in the approaches they consider and of the approaches themselves. The authors analyze different approaches to quantum gravity and arrive at the conclusion that:

> [...] after examining the most popular approaches to QG, we do not find the conceptual discrepancy between GR and the various approaches of QG to be substantive enough to prevent, in principle, a reduction of GR to QG. Insofar as QG embeds a local distinction between two distinct and local quasi-spatial and quasi-temporal structures, this ensures — at least in principle — the possibility to recover the familiar space–time split in relativistic physics from these local separations in the fundamental structure. (Le Bihan & Linnemann, 2019, p. 120)

From the spacetime functionalist perspective,[6] one can read Le Bihan and Linnemann as claiming that there is some distinction between quasi-spatial and quasi-temporal structures in our theories of quantum gravity that allows for fulfilling the spacetime functions needed to recover the spacetime appearance. However, I will argue that one can doubt that this split between quasi-spatial and quasi-temporal is still there once we consider the quantum and dynamical aspects of the putative theories of quantum gravity. In other words, the conceptual discrepancy is greater than what Le Bihan and Linnemann claim, and than what the analysis of spacetime functionalists suggests. In this sense, functionalism and emergentism in general have to deal with the full theories and their dynamical and quantum aspects in order to claim that one is able to recover spacetime from some fundamentally non-spatiotemporal reality. Beyond Le Bihan and Linnemann, spacetime emergentists in general dedicate little space to discussing the conceptual issues associated with the quantum dynamics of quantum gravity. Huggett and Wüthrich (2025) is probably the most notable recent work in which these issues are treated in more detail, and I will argue that even in this case their analysis can be challenged.

## 3. The standard case studies

In this section I will introduce the standard case studies discussed for supporting spacetime emergence. These examples are some kinematical states in LQG and GFT[7] and the classical version of causal set theory. I will argue that the functionalist has a good case, despite some worries, for claiming that spacetime can be reduced to a classical causal set or for claiming that space can be described by some states in LQG. However, I will argue that these examples are limited, as in both cases one is considering just some classical elements of the theories. What the emergentist is showing is that continuum spacetime can be reduced or approximated by discrete structures, but what would be interesting to study is how spacetime is reducible to quantum structures that would be more radically non-spatiotemporal.

### 3.1. Kinematical Hilbert space of LQG and related approaches

Probably the most discussed case of the application of spacetime functionalism is the spin network Hilbert space,[8] which is the kinematical Hilbert space of LQG. Spin networks appear in the canonical formulation of loop quantum gravity and similar Hilbert spaces also appear in other approaches like spin foam models[9] and group field theory.[10] In particular, the spin network Hilbert space has been used for supporting spacetime functionalism or similar positions in Crowther (2016, 2021), Huggett and Wüthrich (2013, 2025), Lam and Wüthrich (2018, 2021) and Wüthrich (2017, 2019a, 2019b).

Loop quantum gravity was originally obtained by applying canonical quantization methods to a particular formulation of general relativity, the connection formulation.[11] The quantization chosen leads to the LQG Hilbert space which can be characterized by its spin network basis. A state in this basis is given by a graph $\Gamma$ and a series of quantum numbers $j, i$ associated with the nodes and links of this graph. By making use of suitable definitions of the area of a surface and the volume of a region one can show that the quantum counterparts of these quantities constitute well-defined operators. States in the spin network basis are eigenstates of these operators. In particular, the area of any surface is a function of the quantum numbers $j$ associated with the links of $\Gamma$ that cross the surface, and the volume of any region is a function of the quantum numbers $i$ associated with the vertices in that region. For this reason, spin networks naturally lead to a picture in which geometry is quantum, as these geometric observables have discrete spectra and superpositions of spin network states would correspond to something like superpositions of geometries.

These states are sometimes interpreted as representing atoms or chunks of space or as representing abstract combinatorial structures. In any case, they are the basis for the spacetime functionalist arguments that the appearance of a continuum space can be explained from something more fundamental. Functionalists deploy different strategies, such as identifying spin networks with tessellations of space, or studying regions with many nodes and links and arguing that, at that scale, the expected values of the quantum operators for area and volume correspond to the geometric properties of a continuum space. While one could raise some worries about the details of this kind of argument, the emergentist claim that space can be explained out of discrete structures is well-exemplified by the spin network case.

[5] I am introducing this caveat just to notice that there are different perspectives on spacetime that may introduce slight modifications to the list. For instance, a Machian could be uncomfortable with the mention of distances and durations and could rather speak about shapes or relative durations.

[6] In Le Bihan and Linnemann (2019) the authors do not commit themselves to functionalism, but just with spacetime emergentism widely understood. I am presenting the argument here in functional terms for ease of presentation, but it can be understood in broader terms. In any case, independently of whether it is understood functionally or otherwise, the arguments in this paper affect Le Bihan and Linnemann's position.

[7] GFT (group field theory) is an approach to quantum gravity which is closely related to LQG. See Mozota Frauca (2023a, 2024a) for a critical assessment of the approach.

[8] Here I will use 'spin network' to refer both to spin networks and s-knots. 'Spin network' originally referred to a network embedded in a space, while 's-knot' was an equivalence class of spin networks under diffeomorphisms. In the recent literature, both in physics and philosophy, one uses the term 'spin network' to refer to both indistinctly.

[9] See Rovelli and Vidotto (2015).

[10] See Oriti (2016).

[11] See Rovelli (2004) for a detailed introduction to LQG.

The limitation of the emergentist analysis of LQG and related approaches is that this analysis concerns the recovery of space, not spacetime. To recover spacetime, we need to consider the dynamics of the theory and see how these structures, close to discrete spaces, evolve. However, emergentist discussions dedicate little attention to dynamics, and it is precisely in the dynamics of this kind of approach where the greatest challenges for the emergentist lie, as I will argue in Section 4.

### 3.2. Causal sets

Causal set theory is another frequent emergentist case study. A theorem by Malament (1977) inspired this program[12] by showing how the metric in general relativity can be understood as composed of the causal relations between points in spacetime and information about volume. A causal set is a discrete version of spacetime that builds upon this understanding of spacetime. A causal set contains countably many elements instead of the continuum of points that we have in a relativistic spacetime, but it is also endowed with a causal structure that tells us whether two elements are causally connected. Each element of the set is like an atom of spacetime, and they are assigned equal spacetime volume. In this sense, a causal set is built with the same kind of ingredients as spacetime: points/regions, causal relations, and volume. It is therefore no surprise that some causal sets are very good approximations to continuum spacetimes when considered at scales much bigger than the scale of the elements of the set. Causal sets can be argued not to be spatiotemporal, given that some causal sets do not approximate spacetimes. For instance, the set can be structured in a way in which a dimensionality cannot be assigned, or that it varies in different parts of the set. That is, we could have a causal set for which some regions approximate a part of a spacetime of dimension 4, some other regions a part of dimension 3, and some other regions do not approximate a spacetime at all. For this reason one can claim, as done for instance in Lam and Wüthrich (2018, Sect. 4), that causal sets are not spatiotemporal and that spacetime emerges in the cases in which causal sets do look like spacetime.

This kind of emergence is similar to the emergence of space from spin networks. Certainly, discrete structures in the right circumstances can be argued to look like continuum ones. But as in the case of approaches like LQG, the interesting case of emergence would happen when we have not only discrete structures, but quantum and dynamical ones. Causal set theory, as a theory of quantum gravity, is an unfinished theory, given that it has defined some classical structures but it has not built a quantum theory from these structures. The physicists working on this approach have hinted at defining objects like transition amplitudes or propagators for causal sets, but as I will argue in the next section, even if we were able to build these objects, their interpretation is challenging.

[12] See the original article (Bombelli et al., 1987) and the recent review (Surya, 2019).

## 4. The missing piece: the quantum dynamics

The discussion in the previous section agrees with the analysis that the spacetime emergentist makes of certain kinematical or classical structures that we find in theories of quantum gravity: they certainly allow us to recover some spatial or spatiotemporal functions and to claim emergence with some plausibility. However, they do not constitute complete theories, and for claiming that spacetime emerges one needs to consider the dynamical and quantum aspects that a theory of quantum gravity necessarily has. The bad news for the emergentist is that it is precisely the interpretation of the dynamical and quantum aspects of quantum gravity that represents a formidable challenge. In this section I will distinguish between two types of approaches, canonical and covariant, and argue that the ways dynamics have been defined in both cases are problematic, in the sense that there are good reasons to doubt that we have a satisfactory interpretation for them. I will note that the conceptual issues that affect these approaches are quite general, which implies that my criticism will not depend on the specific details of the different approaches under consideration. Instead, the issues I will highlight affect the justification and interpretation of the canonical and covariant formalisms as applied to theories with the symmetries of general relativity and hence they are widespread and would also affect future developments using the same formalisms. This jeopardizes the approaches in general, and if one wants to insist that they are describing some non-spatiotemporal reality, then the great challenge for the emergentist is to find a way to make sense of them and argue that they describe something able to fulfill spacetime roles or out of which spacetime could emerge. I will argue that the emergentist has not achieved this and that the optimism of some emergentists is unwarranted as I do not find it plausible that there is some convincing argument able to solve the conceptual issues highlighted in this section.

### 4.1. Canonical approaches

Canonical approaches to quantum gravity are built by applying the methods of canonical quantization to some version of general relativity, some truncation of it, or some close relative. In order to apply these quantization techniques, one introduces a foliation of spacetime into space and time and, then, one quantizes. It is because of this foliation that Le Bihan and Linnemann claim (Le Bihan & Linnemann, 2019) that there is a space–time split in canonical approaches to quantum gravity, which can be used to argue that the gap between the spacetimeless and spacetime is narrow, and hence that emergence is plausible. However, this is too quick, as what we are interested in is the final product of the quantization procedure, and the problem of time of canonical approaches to quantum gravity raises important worries about whether there is something like time in the resulting theory of quantum gravity. In other words, the fact that there was a foliation involved in the quantization procedure does not entail that there is a space–time split in its outcome. The reason for this is that the quantization of reparametrization invariant systems leads to a trivial dynamical equation and one needs to make sense of the theory, supposing it is possible, by means of some different strategy.

It is precisely because of the problem of time that I believe that canonical approaches to quantum gravity are a challenge for the spacetime functionalist. In this kind of approach one can claim that states are 'timeless' and if the functionalist believes that sense can be made of these theories, they need to show how something in these timeless states can play the roles of time, i.e., how a notion of evolution can be recovered, how we can have something like a partial order of events, and so on.

I refer the reader to the literature[13] for reviews and discussions of the problem of time, while here I will just give a brief conceptual introduction. General relativity is a theory with reparametrization invariance, and as such it is treated in a way similar to gauge theories in the Hamiltonian formalism. When quantizing the theory this means that physical states need to satisfy certain constraints and this introduces a distinction between a kinematical Hilbert space in which these constraints have not been imposed and the physical Hilbert space that is obtained by imposing them. This is just analogous to the case in electromagnetism: states in the kinematical Hilbert space can represent any configuration of the electromagnetic field at a time, but only states in the physical Hilbert space represent configurations that are physical in the sense of satisfying Gauss law.[14] The problem of time arises for the case of reparametrization invariant models because in these models the Hamiltonian of the system is a combination of constraints and therefore its action on physical states vanishes. This means that when formulating the Schrödinger equation of the system it is trivial, and physical states do not evolve with respect to the time parameter $\tau$.[15]

[13] In particular, I recommend the two classical reviews (Isham, 1993; Kuchař, 1992), the book (Anderson, 2017) and the more recent critical analysis in Mozota Frauca (2023b, 2024c).

This could mean that something has gone badly with the quantization procedure, but the most extended view in the quantum gravity community is that it is not the case. Instead, what they propose is that states in the physical Hilbert space are all one needs to have a theory of quantum gravity. The challenge for the emergentist (or perhaps for the quantum gravity theorist) is to show how these states describe some timeless or spacetimeless entities and how spacetime could emerge from them. There are several ways in the literature in which it has been argued that one can make sense of this sort of theory without time, but I will argue that they can be challenged and that it seems implausible that spacetime functionalism can solve this issue.

I refer the reader to Mozota Frauca (2023b) for a thorough and critical view that analyzes the different strategies for dealing with the problem of time. Here I will agree with the analysis in there and I will give a summarized version of the general argument in Mozota Frauca (2023b) showing that these strategies are inspired by the case of deparametrizable models, while general relativity is non-deparametrizable, which means that these strategies likely fail for canonical approaches to quantum gravity. A deparametrizable model is a model which is defined on some extended configuration space and time is one of the variables in this space,[16] and this allows for some sort or another of deparametrization which solves the problem. For non-deparametrizable models like general relativity one cannot identify a time variable in the configuration or phase space of the theory, which seems to render the strategies that worked for the deparametrizable models inapplicable.

To see this, consider the case of a reparametrization invariant model for a Newtonian particle, which has been widely discussed in the quantum gravity literature. For this model one has a problem of time in the sense that the evolution with respect to the arbitrary time parameter $\tau$ is trivial, but the problem is solved once one interprets the states $\psi(x, t)$ in the physical Hilbert space not as states to be evolved but as states that already describe an evolution, i.e., once one interprets them as states for the position of the particle that evolve in time $t$. Different resolutions of the problem of time are more or less sophisticated, but for this model they work in similar ways, which involves deparametrizing and identifying $t$ as a time variable.

For non-deparametrizable models one can argue that this is not the case. For instance, take the model studied in Mozota Frauca (2023b), which describes a system of two harmonic oscillators using its Jacobi action.[17] As in the deparametrizable case, states become independent of the time parameter $\tau$. The difference now is that the states $\psi(x, y)$ in the physical Hilbert space do not depend on any variable that can be interpreted as a time variable. Indeed, $x$ and $y$ represent the position of each of the oscillators, which clearly has a physical meaning different from being time. It therefore seems wrong to take one of the two and interpret it as a time variable as one would if one followed one of the strategies that work for the deparametrizable models. That is, it seems wrong to read $\psi(x, y)$ as being a wavefunction in the configuration space of $x$ that evolves with respect to $y$ (or vice versa). Among other reasons, in the classical theory $y$ is oscillating, which means that a value of $y$ does not uniquely determine a moment of time and also that the values of $y$ are bounded. This is in opposition with what we would be claiming if we claimed that $y$ in $\psi(x, y)$ represents time, as it would be taking values in the whole real line and each value of $y$ would be associated with just one instant of time.

By deparametrizing (in one way or another), one is not finding that there are some functions of time that are played by some other entities in a spacetime-less world. On the contrary, one is forcing these functions into them in a problematic way. The parameter $\tau$ encoded an order relation between configurations of the system. While it is true that for a single oscillation we can use $x$ or $y$ as a to keep track of this order relation, when we consider more than one oscillation of the system, this no longer applies. That is, when we are restricted to one oscillation, having $x_1 < x_2$ can tell us that the first configuration happens before the second, but when we consider the whole evolution of the system $x_1 < x_2$ stops giving us information about the order in which the configuration happen, given that $x_1$ obtains before $x_2$ in the same oscillation, but after it if we consider a later oscillation. Forcing a deparametrization on a non-deparametrizable model is to misidentify time. If the quantum theory is to recover the classical theory or something close to it as some sort of limit, it should be able to explain the complicated order relation encoded by $\tau$ and this clearly cannot be done by interpreting $x$ or $y$ as giving the temporal order structure.

Different strategies for solving the problem of time follow different routes, but they share that they are motivated by the deparametrizable case and at the end of the day apply some sort of deparametrization or another. Again, I refer the reader to Mozota Frauca (2023b) for a discussion of this point. As said above, in the works of the spacetime emergentists sometimes little discussion is found about the problem of time, how time disappears, and the way in which it is supposed to be recoverable through functionalism or similar strategies. Huggett and Wüthrich are probably the spacetime emergentists who have discussed most the problem of time (Huggett et al., 2013; Huggett & Wüthrich, 2018, 2025). They claim that the problem of time makes it the case that time and change seem to be missing in quantum gravity, and when discussing how to recover time their discussion is brief and relies on deparametrization or citing Rovelli's partial observables view. In their latest book, Huggett and Wüthrich (2025) they also mention that moving to the covariant formalism may help dealing with the problem of time, and I will discuss this in the next subsection.

Before this, let me briefly discuss Rovelli's partial observable view (Rovelli, 2002, 2004),[18] given that it has a big influence in the emergentist literature. This is the view that in diffeomorphism-invariant or reparametrization invariant models, all the physical content of the theory lies in 'correlations between partial observables'. A partial observable would be something we could measure, and is typically repre-sented by a phase space coordinate. Real measurements and predictions would be represented by 'complete observables', for which more than one partial observable needs to be measured. For instance, when we say that we measure the position of a single particle, we measure it at a time, so there are two variables at play, the position, and the time. In this sense, the complete observable is, using Rovelli's notation, $X_T$. $X_T$ can be thought of as representing $X$ as a function of $T$, and this is what Rovelli calls a correlation between partial observables.

[14] For completeness, let me say that not every constraint in the Hamiltonian formalism has such a nice interpretation in the original Lagrangian theory, given that some just fix some momenta to a certain value, with no further physical interpretation.

[15] The same argument works in the Heisenberg picture or in any equivalent representation of the dynamics.

[16] For field theories the same discussion applies but one may need to speak about spacetime coordinates instead of a time variable.

[17] The relevance of this kind of action for the discussion of the problem has been highlighted by Barbour (1994a, 1994b) and Gryb (2010).

[18] See also Rovelli (2011), Rovelli and Vidotto (2015, 2022) and the criticisms in Mozota Frauca (2024b, 2024d, 2025a) and Thébault (2012).

It is not difficult to see that this is just an example of a deparametrization. It works fine for deparametrizable models, but it shows conceptual shortcomings if we try to apply it to non-deparametrizable models. In the example of the two harmonic oscillators, the two partial observables would be the positions of the two harmonic oscillators, and the complete observables correlations of the form $X_Y$ or $Y_X$. This is of course problematic, as neither $x$ nor $y$ are monotonic in $\tau$, $X_Y$ and $Y_X$ are not well defined as functions, as they are multivalued. And again, $\tau$ encoded a temporal order that is not represented by $X_Y$ or $Y_X$ and that we would like to preserve. In the quantum version of the theory, Rovelli and collaborators seem fine with defining expectation values for $X_Y$ or $Y_X$ (or formulating an equivalent version of this), but this again defines an expectation value for $x$ at some fixed value of $y$ or the other way around, when what we ought to have is at least a formalism able to make different predictions for each time the second oscillator is at $y$, if we are to recover the classical theory and spatiotemporal structure, even if approximately.

The shortcomings of the partial observable view are well exemplified by its application to cosmological settings. For instance at Rovelli (2004, p. 298), Rovelli claims that while in cosmological models we may have a description of how the scale factor of the universe $a$ and a scalar field $\phi$ evolve in time $t$, we ought to express the evolution of $a$ as a function of $\phi$ or of $\phi$ as a function of $a$.[19] Rovelli is therefore inviting us to throw away the physical content encoded in $t$, but this just falls prey of the conceptual worries highlighted above: $a$ and $\phi$ are not time variables (even if they can be monotonic in some models) and treating one of them as a time variable and the other one as a 'quantum' one seems wrong. Moreover, $t$ is necessary for making important predictions in cosmology such as all the predictions related to duration. For instance, the claims that the universe is 13.8 billion years old or that the duration of the early phases of our universe explain the abundance of Helium are empirical claims that the partial observable view seems to overlook. In other words, the claim that it is just correlations between $a$ and $\phi$ that are physically meaningful seems to be begging the question, and the spacetime emergentist should ask for a way of recovering time. For a discussion of the problem of time in cosmological settings, I refer the reader to Mozota Frauca (2025b, 2026), where similar worries are raised.

Alternatively, one can read some works in canonical quantum gravity as accepting that interpretations based on deparametrization are not the way to go and that some different, possibly more radically new interpretation is needed. However, as argued in Mozota Frauca (2023b), these interpretations need to be spelled out and it is not clear that a functionalist argument is able to recover the functions of time from such 'timeless' states. For this reason, I believe the emergentist story regarding canonical approaches like LQG as discussed in Section 3.1 is incomplete and these issues regarding the problem of time need to be addressed in order to present an argument supporting that LQG, or some other canonical approach, describes a non-spatiotemporal reality which nevertheless can look spatiotemporal. Until then, skepticism about canonical approaches and the claim that they describe an emergent spacetime seems a reasonable position.

[19] Spacetime emergentists also fall into this kind of deparametrization in Huggett and Wüthrich (2018).

### 4.2. Covariant approaches

Let me now move to the other main family of approaches to quantum gravity: covariant approaches. These approaches are built by means of path integrals or sum over histories techniques and they define different structures from the ones defined by canonical quantization. However, they also face significant interpretational challenges which jeopardize the claims that the spacetime emergentist would like to make about them, as I will argue in this section.

Let me start by noticing that string theory would be naturally included in this category, but the analysis will be focused on approaches that attempt to quantize general relativity or some close version of it. Some authors have analyzed the way in which spacetime appears in string theory and have also argued for an emergentist position in this context. While I think that some of my criticisms may be translatable to the string theory context, it is beyond the scope of this article to discuss this theory, and I refer the reader to Huggett and Vistarini (2015), Huggett and Wüthrich (2013, 2025) and Huggett et al. (2013) for discussions of the role of spacetime in string theory.

Having said this, we can go back to the analysis in Le Bihan and Linnemann (2019) and what they say about these approaches. For instance, for causal set theory they claim that the causal structure of the causal set marks a distinction between something quasi-temporal and something quasi-spatial. Similarly, they take the structures of other approaches like causal dynamical triangulations or spin foam models to be encoding some similar distinction. However, concluding from this that there is such a distinction in the theories of quantum gravity they define is too quick, as even if structures like causal sets, spin foams, or triangulations have a time–space split, the theory considers sums of several of these structures and it is wrong to infer something about what the theory is about just by looking at the properties of each term in the sum. This would be analogous to saying that as one can express the quantum theory of a single particle using a path integral formalism in which one sums over continuous trajectories one can infer that the particle follows a continuous trajectory. This conclusion can be contested and has been contested, and this should make us cautious when interpreting theories of quantum gravity.

For instance, in the case of causal set theory, each term in the sum represents a well-defined causal structure and for each pair of events we can tell whether they are timelike or spacelike. However, when we consider a sum of terms representing different causal sets, we find different causal structures: a pair of events can be timelike in one set and spacelike in another.[20] Perhaps in this case it is better to say that the theory describes something with an indefinite causal structure or refrain from making claims about it and whether it has a time–space split. This is just as in the case of the path integral for a single particle, where one should be careful about the inferences one makes about it following a well-defined or continuous trajectory.

The interpretation of covariant approaches to quantum gravity is challenging and this difficulty, as far as I know, has not been fully addressed by the emergentist camp. For standard quantum mechanics or quantum field theory the interpretation of the theory is generally considered independently of which quantization techniques were used for formulating the theory and there are a few well-established interpretations of the formalism. However, when we move to covariant approaches to quantum gravity we find that the standard structures of quantum mechanics are not defined in these approaches and this complicates interpreting the formalism. That is, if we do not have a time-evolving quantum state we cannot apply interpretations like Everett quantum mechanics, Bohmian mechanics, or collapse models.

[20] Identification of elements in different causal sets may be tricky or not even possible, but let me keep these worries aside, as they do not affect my argument that the causal structure is different.

The most common structure defined by covariant approaches is the 'propagator', also known as the 'transition amplitude' of the theory, which is a mathematical object of the form $K(\text{fin}, \text{in})$, or $K(\text{bound})$, where 'fin', 'in' and 'bound' represent a state at a final moment, at an initial moment or at a boundary of a spacetime region. These 'propagators' are defined by means of some sort of path integral, just as it can be done in ordinary quantum mechanics. However, there is an important difference. These objects are not genuine propagators, in the sense that they are not a representation of a time-evolution operator. That is, while the propagator of quantum mechanics is an object with a temporal dependence (it is of the form $K(\text{fin}, t_{\text{fin}};\text{in}, t_{\text{in}})$) that can be used to determine the quantum state of a system at any time when the state at some initial moment is known, the 'propagators' defined in the quantum gravity context lack time dependence and cannot be used for recover a time-evolving state. We see then how time-dependence also disappears in the covariant formalism, and hence that there is a covariant version of the problem of time.

As in the canonical case, the distinction between deparametrizable and non deparametrizable models is very relevant here. For both kind of models the dependence of the arbitrary parameter $\tau$ is dropped when building the 'propagators'. But just as in the canonical case, the fact that $t$ is a variable in the configuration state of deparametrizable models, allows for avoiding the problem. More precisely, the propagator defined is not a propagator in $\tau$-time, but it has the form of a propagator in $t$-time and standard quantum mechanics can be recovered. In the non-deparametrizable case, we are not this lucky and we have no time variable to perform a deparametrization. And just as in the canonical case, forcing a deparametrization and interpreting the 'propagator' as a propagator in $x$-time or $y$-time is conceptually mistaken and unable to recover the classical behavior of our systems, as discussed above. In this sense, the problem of time is also present in the covariant formalism, which means that it also represents a challenge for the interpretation of the theory. I believe that the emergentist should consider it carefully if they want to argue that the formalism is describing a world with no time at the fundamental level but in which time can be recovered in some functionalist or similar way.

In the covariant quantum gravity literature, 'propagators' are mostly interpreted by following two strategies. The first one is to relate them with canonical approaches to quantum gravity. Indeed, in some approaches the 'propagator' aims to be a representation of the projector which maps states from a kinematical Hilbert space to a physical Hilbert space. In the case of spin foam models, the 'propagator' would represent the projector from the spin network Hilbert space to the physical Hilbert space of canonical LQG.[21] For this reason, we could take the definitions of the 'propagators' of different covariant approaches just as an alternative way of defining the physical Hilbert space of a canonical quantization. But then, if the covariant formalism is just a rewriting of the canonical one, my arguments in the previous section directly apply. The fact that the 'propagator' is defined as a sum of 'spacetime-like' structures like spin foams does not make the interpretation of states in the physical Hilbert space any easier, and it remains a challenge for the interpretation of the theory of quantum gravity and for the case of the spacetime emergentist.

The alternative is to interpret the 'propagator' as encoding 'the probability of finding a final state/configuration/geometry, given that an initial one obtained'. However, making a sensible interpretation of this seems difficult, especially from the realist perspective of the functionalist project. From the claim that $K(\text{fin},\text{in})$ defines a probability it is hard to answer questions about how is the world like according to this theory. But it is even worse, as even from an operationalist perspective it is difficult to make sense of initial and final measurements/observations/preparations if they are not embedded in some spatiotemporal structure. Intuitively, one would expect the 'propagator' to carry some temporal dependence in such a way that the probability of finding a given state depends, in some way or another, on the amount of time that happens until the measurement.

Similarly, we can consider again the case of the model of two harmonic oscillators. The 'propagator' for this theory is an object of the form $K(x, y; x', y')$. The probabilistic interpretation of it would read it as encoding the probability of finding the oscillators with positions $x, y$ if we know that they had positions $x', y'$. The lack of temporal dependence is again shocking: we would naively think that the probability distribution should depend on how much we wait to make the measurement. If in the classical theory we had sequences of configurations, our intuition for the quantum theory is that we should have sequences of probability distributions. For instance, we would like to be able to claim that if we start with a distribution of probabilities peaked around a classical configuration, it should evolve and become a peaked distribution around a configuration obtained by evolving the original distribution. This of course means that the probability should change, but change is missing from this formalism.

In the literature we find again that many authors get very close to deparametrize the model. In the partial observable view introduced above, sometimes objects like $K(x, y; x', y')$ are interpreted as a probability distribution for $x$ when $y$ takes some given value. Even if at first this may sound appealing, it is easy to see that it falls prey to the same conceptual issues as in the canonical case. For instance, if the model is to recover the classical behavior in some approximate manner we would like to distinguish the probability distribution for $x$ when $y$ is at position $y_0$ at its first oscillation, when it is at its second, and so on. Even if the partial observable view is right in claiming that we many times are interested in predictions of the form 'the value of one quantity when another takes some value', this does not capture the temporal structure of our models, which is more complicated. Again, framing dynamics in terms of correlations misses part of the point and is unable to solve the conceptual problems that arise when the formalism is truly timeless.

This example also shows that the operationalist framing is quite obscure and not very helpful when trying to figure out what the theory is about, whether it is timeless or not, and how to recover time. In the classical theory, we have two harmonic oscillators moving in space, and if we want to think about observers we could have them as external systems. When moving to the standard, timeful version of quantum mechanics, from the operationalist view one may not want to say anything about what is happening to the system, but at some times it interacts with the 'observer' and that quantum mechanics gives us some rules for predicting the probabilities for the different possible outcomes. Operationalism is not committed to saying anything about the system in between measurements, but notice that it is framed in a spatiotemporal context: the system is in spacetime and measurements are interactions that happen at some spacetime region. There is a common spacetime structure that operationalism needs. At least, the time parameter of timeful quantum mechanics represents the time that elapses both for the system and for the observer.

When we move to the case in which there is no time in the formalism, just appealing to operationalism as a way of solving interpretational issues does not seem promising. Now the quantum formalism lacks any time parameter, while the standard way of thinking about observers, observations and measurements, still depends on plenty of temporal intuitions. If we say that an observer performs a measurement after a previous observation it seems that we need to introduce a minimal temporal structure. Here I think the quantum gravity theorist or the spacetime emergentist need to think about how to understand this claim. One way of doing it would be that timeless quantum systems, as the double harmonic oscillator or whatever our quantum gravity model aims to describe, are timeless, while 'the world of observers' would be timeful. Somehow, we would need to embed the quantum system in spacetime, in order to relate it to what observers would observe. But here again we see that one would expect the probabilities to depend on the way the system is embedded. That is, the initial and final measurements will be located at some time $t$ from each other according to the external temporal structure and we again would expect that the probability should depend on this $t$. Again, if that is not the case, the classical limit of the theory does not look available. Let me mention that this operational strategy has been proposed in Rovelli and Vidotto (2015) and presented in Huggett and Wüthrich (2025) for the case of quantum gravity and as a way of dealing with the problem of time. That is, it has proposed that some 'propagators' define probabilities for some measurements of 'quantum spacetime' from outside, which would be like performing a measurement at the boundary between a classical spacetime and the 'quantum' region. Again, we are embedding the configurations of the propagators in spacetime, but the covariant version of the problem of time means that these 'propagators' are independent of how we embed the system in spacetime, i.e., which boundaries we choose. And just as in the case of the double harmonic oscillator, if we are to recover general relativity, we better have probabilities depending on when and where we place the initial and final boundaries.

[21] See the discussion in Rovelli and Vidotto (2015).

The alternative of interpreting the temporal-seeming claims of the operational talk as requiring some temporal structure on the observers' side is to read this sort of claim in some different, sui generis sense. Surely, if we are thinking about something like quantum spacetime, it is likely that we should revise our notions of interaction or measurement, but this is an important work that needs to be done. To the best of my knowledge, such a revision has not been performed, and hence the claim that the 'propagator' represents a probability remains quite obscure, and the way in which from that claim one could recover spacetime is not straightforward at all. For this reason, I believe that it is sensible to press on the spacetime emergentist and quantum gravity theorist and ask them to provide an account able to overcome these worries.

Finally, let me also mention that some approaches like causal dynamical triangulations have focused more on defining different mathematical structures like partition functions. However, from a conceptual point of view the situation is similar to the situation with 'propagators': we have more or less rigorous definitions of some mathematical objects but no clear interpretation for them. The standard connexions between partition functions, propagators and dynamics are not straightforward when mathematical objects lack the spacetime dependence they had in the timeful context. That is, even if the objects we are defining have a resemblance (and the same names) to objects in quantum mechanics or quantum field theories, if we can connect them with a canonical formalism it is the timeless one, and if we opt to interpret them independently, the spacetime emergentist should give us an account of how this interpretations is to be made. Just as for propagators, partition functions are also related to probabilities in the timeful context, but their interpretation becomes obscure when the formalism becomes timeless. The challenge for the emergentist is to interpret which non-spatiotemporal reality is supposedly described by such objects and the way they could constitute an emergent spacetime.

To insist, when we consider not just a structure like a causal set, a spin foam, or a simplicial decomposition, but some sort of sum over different such structures, the interpretation of what the model is about becomes challenging, if possible at all. As noted above, this is more complicated than the case of standard quantum mechanics, where at least realist interpretations could offer some way of bridging between the formalism and the real world. Here we are many times left with the claim that these theories define some probabilities, but it is hard to make sense of this sort of claim. It is in this sense that the emergentist picture is also incomplete for covariant approaches to quantum gravity, as it is not clear what the fundamental entities of these approaches are and the way space and time could emerge from them.

## 5. Conclusion

In this paper I have argued that the emergentist project is a promising way of understanding how a theory that could be fundamentally non-spatiotemporal can relate to our spatiotemporal-looking world. However, I have argued that the most challenging part for the understanding of our theories of quantum gravity is that they are quantum, and that they are quantum in a way that differs from the standard dynamics of quantum mechanics or quantum field theory. I have noted that the different approaches to quantum gravity, both canonical and covariant, can be questioned precisely for this reason, and that a good argument can be made for claiming that we lack a satisfactory interpretation for them. It is therefore a challenge for the spacetime emergentist to complete the picture they want to present and consider full theories and not just some kinematical or classical aspects of the theories, as it is precisely in the quantum and dynamical aspects of these approaches where the greatest conceptual difficulties lie.

The arguments in this paper can be seen as a challenge to the viability of several approaches to quantum gravity. Indeed, the arguments provided are quite general and affect two big families of approaches, in particular those which have been used to argue for a fundamentally spacetimeless reality. The arguments in Section 4 support the conclusion that the quantum and dynamical aspects of not only current approaches to quantum gravity, but of the general formal structures accepted by the quantum gravity community, make them unsatisfactory and not constituting meaningful physical theories.

One of the goals of spacetime emergentism was precisely to help make sense of current approaches to quantum gravity, and authors like Linnemann and Le Bihan argued that it should be able to bridge the gap between them and general relativity. According to the analysis here, functionalism is of no help in solving the general conceptual issues I have highlighted, and therefore functionalism fails in reaching this goal.

Finally, even if I have not argued against the general case that a spacetimeless theory out of which spacetime could be recovered, I believe that the case for spacetime functionalism (in the context of quantum gravity) or emergentism is greatly weakened by the fact that the discussions by the emergentists are many times reduced to just a few classical features or to discrete spacetimes and not to something truly non-spatiotemporal and quantum.

## Acknowledgments

I want to thank the philosophy of physics reading group at UCSD, the Proteus group, Carl Hoefer, Brian Pitts and Daniele Oriti for their comments and discussions. This research is part of the Proteus project that has received funding from the European Research Council (ERC) under the Horizon 2020 research and innovation programme (Grant agreement No. 758145) and of the project CHRONOS (PID2019-108762GB-I00) of the Spanish Ministry of Science and Innovation.

## Data availability

No data was used for the research described in the article.

## References

Anderson, E. (2017). Fundamental theories of physics. In *Fundamental theories of physics publication*: *vol. 190*, *The problem of time*. Cham: Springer International Publishing, http://dx.doi.org/10.1007/978-3-319-58848-3, URL: http://link.springer.com/10.1007/978-3-319-58848-3.

Barbour, J. B. (1994a). The timelessness of quantum gravity: I. The evidence from the classical theory. *Classical and Quantum Gravity*, *11*(12), 2853. http://dx.doi.org/10.1088/0264-9381/11/12/005, URL: https://iopscience.iop.org/article/10.1088/0264-9381/11/12/005.

Barbour, J. B. (1994b). The timelessness of quantum gravity: II. The appearance of dynamics in static configurations. *Classical and Quantum Gravity*, *11*(12), 2875. http://dx.doi.org/10.1088/0264-9381/11/12/006, URL: https://iopscience.iop.org/article/10.1088/0264-9381/11/12/006.

Bombelli, L., Lee, J., Meyer, D., & Sorkin, R. D. (1987). Space-time as a causal set. *Physical Review Letters*, *59*(5), 521–524. http://dx.doi.org/10.1103/PhysRevLett.59.521, URL: https://journals.aps.org/prl/abstract/10.1103/PhysRevLett.59.521.

Crowther, K. (2016). *Effective spacetime: Understanding emergence in effective field theory and quantum gravity*. Springer International Publishing, http://dx.doi.org/10.1007/978-3-319-39508-1, Publication Title: Effective Spacetime: Understanding Emergence in Effective Field Theory and Quantum Gravity.

Crowther, K. (2021). As below, so before: 'synchronic' and 'diachronic' conceptions of spacetime emergence. *Synthese*, *198*(8), 7279–7307. http://dx.doi.org/10.1007/s11229-019-02521-1, arXiv:1912.12065.

Esfeld, M. (2021). Against the disappearance of spacetime in quantum gravity. *Synthese*, *199*(S2), 355–369. http://dx.doi.org/10.1007/s11229-019-02168-y, URL: http://link.springer.com/10.1007/s11229-019-02168-y.

Gryb, S. (2010). Jacobi's principle and the disappearance of time. *Physical Review D*, *81*(4), Article 044035. http://dx.doi.org/10.1103/PhysRevD.81.044035, URL: https://journals.aps.org/prd/abstract/10.1103/PhysRevD.81.044035.

Huggett, N., & Vistarini, T. (2015). Deriving general relativity from string theory. *Philosophy of Science*, *82*(5), 1163–1174. http://dx.doi.org/10.1086/683448, URL: https://www.cambridge.org/core/journals/philosophy-of-science/article/abs/deriving-general-relativity-from-string-theory/E113D9197476E7B99360D173E6245FE3.

Huggett, N., Vistarini, T., & Wüthrich, C. (2013). Time in Quantum Gravity. *A Companion to the Philosophy of Time*, 242–261. http://dx.doi.org/10.1002/9781118522097.ch15, arXiv:1207.1635 ISBN: 9780470658819.

Huggett, N., & Wüthrich, C. (2013). Emergent spacetime and empirical (in)coherence. *Studies in History and Philosophy of Science Part B - Studies in History and Philosophy of Modern Physics*, *44*(3), 276–285. http://dx.doi.org/10.1016/j.shpsb.2012.11.003, arXiv:1206.6290.

Huggett, N., & Wüthrich, C. (2018). The (A)temporal emergence of spacetime. *Philosophy of Science*, *85*(5), 1190–1203. http://dx.doi.org/10.1086/699723, URL: https://www.journals.uchicago.edu/doi/10.1086/699723.

Huggett, N., & Wüthrich, C. (2025). *Out of nowhere: The emergence of spacetime in theories of quantum gravity*. Oxford, New York: Oxford University Press.

Isham, C. J. (1993). Canonical quantum gravity and the problem of time. *Integrable Systems, Quantum Groups, and Quantum Field Theories*, 157–287. http://dx.doi.org/10.1007/978-94-011-1980-1_6, arXiv:gr-qc/9210011.

Knox, E. (2013). Effective spacetime geometry. *Studies in History and Philosophy of Science Part B - Studies in History and Philosophy of Modern Physics*, *44*(3), 346–356. http://dx.doi.org/10.1016/j.shpsb.2013.04.002.

Knox, E. (2014). Spacetime structuralism or spacetime functionalism? *Manuscript*.

Kuchař, K. V. (1992). Time and interpretations of quantum gravity. In G. Kunstatter, D. Vincent, & J. Williams (Eds.), *Proceedings of the 4th Canadian conference on general relativity and relativistic astrophysics*. Singapore: World Scientific Publishing Company, http://dx.doi.org/10.1142/S0218271811019347.

Lam, V., & Esfeld, M. (2013). A dilemma for the emergence of spacetime in canonical quantum gravity. *Studies in History and Philosophy of Science Part B - Studies in History and Philosophy of Modern Physics*, *44*(3), 286–293. http://dx.doi.org/10.1016/j.shpsb.2012.03.003.

Lam, V., & Wüthrich, C. (2018). Spacetime is as spacetime does. *Studies in History and Philosophy of Science. Part B. Studies in History and Philosophy of Modern Physics*, *64*, 39–51. http://dx.doi.org/10.1016/j.shpsb.2018.04.003.

Lam, V., & Wüthrich, C. (2021). Spacetime functionalism from a realist perspective. *Synthese*, *199*, 335–353. http://dx.doi.org/10.1007/s11229-020-02642-y, arXiv:2003.10172.

Le Bihan, B. (2018a). Priority monism beyond spacetime. *Metaphysica*, *19*(1), 95–111. http://dx.doi.org/10.1515/mp-2018-0005.

Le Bihan, B. (2018b). Space emergence in contemporary physics: Why we do not need fundamentality, layers of reality and emergence. *Disputatio*, *10*(49), 71–95. http://dx.doi.org/10.2478/DISP-2018-0004.

Le Bihan, B. (2021). Spacetime emergence in quantum gravity: functionalism and the hard problem. *Synthese*, *199*(2), 371–393. http://dx.doi.org/10.1007/s11229-019-02449-6.

Le Bihan, B., & Linnemann, N. S. (2019). Have we lost spacetime on the way? Narrowing the gap between general relativity and quantum gravity. *Studies in History and Philosophy of Science Part B - Studies in History and Philosophy of Modern Physics*, *65*, 112–121. http://dx.doi.org/10.1016/j.shpsb.2018.10.010.

Malament, D. B. (1977). The class of continuous timelike curves determines the topology of spacetime. *Journal of Mathematical Physics*, *18*(7), 1399–1404. http://dx.doi.org/10.1063/1.523436, URL: https://aip.scitation.org/doi/abs/10.1063/1.523436.

Martens, N. C. (2019). The metaphysics of emergent spacetime theories. *Philosophy Compass*, *14*(7), 1–15. http://dx.doi.org/10.1111/phc3.12596.

Mozota Frauca, Á. (2023a). Geometrogenesis in GFT: an analysis. *Philosophy of Physics*, *1*(1), URL: http://arxiv.org/abs/2307.11805. arXiv:2307.11805.

Mozota Frauca, Á. (2023b). Reassessing the problem of time of quantum gravity. *General Relativity and Gravitation*, *55*(1), 21. http://dx.doi.org/10.1007/s10714-023-03067-x, URL: https://arxiv.org/abs/2301.07973v1. arXiv:2301.07973.

Mozota Frauca, Á. (2024a). Foundational issues in group field theory. *Foundations of Physics*, *54*(3), 33. http://dx.doi.org/10.1007/s10701-024-00763-9.

Mozota Frauca, Á. (2024b). GPS observables in Newtonian spacetime or why we do not need 'physical' coordinate systems. *European Journal for Philosophy of Science*, *14*(4), 51. http://dx.doi.org/10.1007/s13194-024-00611-7.

Mozota Frauca, Á. (2024c). The problem of time for non-deparametrizable models and quantum gravity. In F. Bianchini, V. Fano, & P. Graziani (Eds.), *The SILFS series: vol. 4, Current topics in logic and the philosophy of science. papers from SILFS 2022 postgraduate conference*. College Publications, URL: https://philsci-archive.pitt.edu/24052/.

Mozota Frauca, Á. (2024d). Time is order. In S. De Bianchi, M. Forgione, & L. Marongiu (Eds.), *Time and timelessness in fundamental physics and cosmology: historical, philosophical, and mathematical perspectives* (pp. 49–67). Cham: Springer Nature Switzerland, http://dx.doi.org/10.1007/978-3-031-61860-4_4.

Mozota Frauca, Á. (2025a). Against radical relationalism: in defense of the ordinal structure of time. *Foundations of Physics*, *55*(3), 37. http://dx.doi.org/10.1007/s10701-025-00850-5.

Mozota Frauca, Á. (2025b). Quantum cosmology and the age of the universe. *Journal of Physics: Conference Series*, *2948*(1), Article 012008. http://dx.doi.org/10.1088/1742-6596/2948/1/012008.

Mozota Frauca, Á. (2026). Does quantum cosmology predict the age of the universe? *Journal for General Philosophy of Science*, http://dx.doi.org/10.1007/s10838-025-09754-4.

Oriti, D. (2016). Group field theory as the second quantization of loop quantum gravity. *Classical and Quantum Gravity*, *33*(8), 1–24. http://dx.doi.org/10.1088/0264-9381/33/8/085005, arXiv:1310.7786.

Oriti, D. (2017). The universe as a quantum gravity condensate. *Comptes Rendus Physique*, *18*(3–4), 235–245. http://dx.doi.org/10.1016/j.crhy.2017.02.003, arXiv:1612.09521.

Rovelli, C. (2002). Partial observables. *Physical Review D*, *65*(12), Article 124013. http://dx.doi.org/10.1103/PhysRevD.65.124013, URL: https://journals.aps.org/prd/abstract/10.1103/PhysRevD.65.124013.

Rovelli, C. (2004). *Quantum gravity*. Cambridge: Cambridge University Press, http://dx.doi.org/10.1017/CBO9780511755804, URL: https://www.cambridge.org/core/product/identifier/9780511755804/type/book.

Rovelli, C. (2011). "Forget time" essay written for the FQXi contest on the nature of time. *Foundations of Physics*, *41*(9), 1475–1490. http://dx.doi.org/10.1007/s10701-011-9561-4, URL: https://link.springer.com/article/10.1007/s10701-011-9561-4.

Rovelli, C., & Vidotto, F. (2015). *Covariant loop quantum gravity: An elementary introduction to quantum gravity and spinfoam theory*. Cambridge: Cambridge University Press, http://dx.doi.org/10.1017/CBO9781107706910, Publication Title: Covariant Loop Quantum Gravity: An Elementary Introduction to Quantum Gravity and Spinfoam Theory.

Rovelli, C., & Vidotto, F. (2022). Philosophical foundations of loop quantum gravity. URL: https://arxiv.org/abs/2211.06718v2.

Surya, S. (2019). The causal set approach to quantum gravity. *Living Reviews in Relativity*, *22*(1), 1–75. http://dx.doi.org/10.1007/s41114-019-0023-1, arXiv:1903.11544.

Thébault, K. P. (2012). Three denials of time in the interpretation of canonical gravity. *Studies in History and Philosophy of Science. Part B. Studies in History and Philosophy of Modern Physics*, *43*(4), 277–294. http://dx.doi.org/10.1016/J.SHPSB.2012.09.001.

Wüthrich, C. (2017). Raiders of the lost spacetime. In *Towards a theory of spacetime theories* (pp. 297–335). New York, NY: Birkhäuser, http://dx.doi.org/10.1007/978-1-4939-3210-8_11, URL: https://link.springer.com/chapter/10.1007/978-1-4939-3210-8_11. arXiv:1405.5552.

Wüthrich, C. (2019a). The emergence of space and time. In *The routledge handbook of emergence* (pp. 315–326). Routledge, http://dx.doi.org/10.4324/9781315675213-26, URL: https://www.taylorfrancis.com/books/9781317381501/chapters/10.4324/9781315675213-26. arXiv:1804.02184.

Wüthrich, C. (2019b). Quantum gravity from general relativity. URL: http://arxiv.org/abs/1902.02099. arXiv:1902.02099.